# Microheater hotspot engineering for repeatable multi-level switching in foundry-processed phase change silicon photonics

Hongyi Sun,[1,2] Chuanyu Lian,[1,2] Francis Vásquez-Aza,[3] Sadra Rahimi Kari,[4] Yi-Siou Huang,[1,2] Alessandro Restelli,[5] Steven A. Vitale,[6] Ichiro Takeuchi,[1] Juejun Hu,[7] Nathan Youngblood,[4] Georges Pavlidis,[3] Carlos A. Ríos Ocampo[1,2,*]

[1]Department of Materials Science & Engineering, University of Maryland, College Park, MD, 20742, USA
[2]Institute for Research in Electronics and Applied Physics, University of Maryland, College Park, MD, 20742, USA
[3]Department of Mechanical Engineering, University of Connecticut, Storrs, CT 06269, USA
[4]Electrical & Computer Engineering Department, The University of Pittsburgh, Pittsburgh, PA 15213, USA
[5]Joint Quantum Institute, Department of Physics, University of Maryland, College Park, MD, 20742, USA
[6]Advanced Materials and Microsystems Group, MIT Lincoln Laboratory, Lexington, MA, 02139, USA
[7]Department of Materials Science & Engineering, MIT, Cambridge, MA, 02139, USA
* corresponding author: riosc@umd.edu

## Abstract

Nonvolatile photonic integrated circuits employing phase change materials have relied either on optical switching mechanisms with precise multi-level control but poor scalability or electrical switching with seamless integration and scalability but mostly limited to a binary response. Recent works have demonstrated electrical multi-level switching; however, they relied on the stochastic nucleation process to achieve partial crystallization with low demonstrated repeatability and cyclability. Here, we re-engineer waveguide-integrated microheaters to achieve precise spatial control of the temperature profile (i.e., hotspot) and, thus, switch deterministic areas of an embedded phase change material cell. We experimentally demonstrate this concept using a variety of foundry-processed doped-silicon microheaters on a silicon-on-insulator platform to trigger multi-step amorphization and reversible switching of $Sb_2Se_3$ and $Ge_2Sb_2Se_4Te$ alloys. We further characterize the response of our microheaters using Transient Thermoreflectance Imaging. Our approach combines the deterministic control resulting from a spatially resolved glassy-crystalline distribution with the scalability of electro-thermal switching devices, thus paving the way to reliable multi-level switching towards robust reprogrammable phase-change photonic devices for analog processing and computing.

# Introduction

Chalcogenide phase change materials (PCMs) are a promising material platform for photonics, given their unique combination of nonvolatility and large refractive index modulation.[1] PCMs such as $Ge_2Sb_2Te_5$ (GST), $Sb_2Se_3$, $Ge_2Sb_2Se_4Te$ (GSST) allow reversible nonvolatile switching between the stable crystalline and amorphous phases. Intermediate mixtures are also possible, thus allowing for a semi-continuous modulation of the refractive index or the extinction coefficients, thereby controlling the phase and amplitude of light. These properties promise devices with zero-static power consumption and ultra-compact form factors, which can complement the ultrafast, broadband, and energy-efficiency characteristics of photonics technology.[2–4] Indeed, since 2010, there has been a rapidly increasing number of studies employing PCMs in nanophotonic applications, namely photonic memories,[5–7] optical computing,[8–11] optical switching,[12–14] metasurfaces,[15–17] etc.

PCM-based photonic devices rely on precisely switching between the amorphous and crystalline states of the PCM, which is achieved utilizing optical or electrical pulses. Optical pulses from within a waveguide[6,18] or a free-space source[19,20] can be used to switch a PCM cell embedded into a photonic integrated circuit (PIC). The PCM directly absorbs energy from pulses to generate heat, leading to near-GHz, low-energy operations and, remarkably, a large number of multi-level states that result from the deterministic control of the amorphous/crystalline spatial distribution in the PCM cell.[21] However, this high efficiency is available only for a few cells due to the slow alignment process of free-space lasers or the complex routing of on-chip pulses. On the other hand, electrical switching relies on the electro-thermal response of microheaters to switch large PCMs since traditional cross-bar devices create filaments that are too small to cause a sizable optical response.[22] Various microheaters, readily scalable to large architectures, have been tested using material platforms including metals,[16,23] graphene,[24,25] doped silicon,[26–30] and transparent conducting oxides.[31,32] From this list, doped-silicon microheaters have demonstrated maximum versatility due to their compatibility with CMOS processes and the ease of back-end-of-the-line (BEOL) integration to silicon-on-insulator (SOI) platforms. Microheaters, however, have relied on geometrical shapes that lead predominantly to flat temperature profiles that trigger a transition in the entire PCM cell. Thus, repeatable multi-level response with electrical switching has proven to be a more significant challenge than optical switching because it relies on a stochastic process: nucleation, which randomizes the intermediate states.[33] To overcome this challenge, researchers have proposed mechanisms to achieve more intermediate levels by modifying the heater,[27,32] but with limited cyclability demonstrated. Moreover, with operating binary PCMs cells, other proposals have shown a series of $N$ devices on a single waveguide that achieve $\sim 2^N$ levels,[34–36] with the disadvantage of increasing the number of electrical contacts and overall footprint.

Here, we experimentally demonstrate novel doped-silicon microheaters to overcome both challenges simultaneously and achieve deterministic multi-level electrical switching of any PCM in a single heater configuration. To do so, we implement an electro-thermal device that enables controllable amorphous/crystalline spatial distributions, and thus, reliable intermediate states. We achieve this by engineering the spatial profile of the microheaters' hotspot using multi-bridge geometries with fixed and varying widths. We demonstrate this approach using two instances of CMOS foundry fabrication combined with BEOL processes, including a zero-change commercial foundry.

## Results

Fig. 1 shows the structures of three different microheater geometries: Type I: bowtie, Type II: five identical 10×2 μm bridges with 2 μm spacing, and Type III: five 5 μm-long bridges with varying widths of 1, 1.5, 3, 1.5, 1 μm and 1 μm separation. We performed Finite-Element Method simulations (FEM) using COMSOL Multiphysics® to obtain the 3D and the surface temperature profiles after applying an electrical pulse to each of the microheaters. Fig. 1(a) shows the simulation results for the commonly used bowtie geometry.[28,37] This microheater reaches a nearly flat temperature profile, meaning the entire PCM cell reaches a uniform crystallization or amorphization if the maximum temperature is above the crystallization temperature ($T_c$) or the melting temperature ($T_m$), respectively. This geometry relies on the stochastic nucleation process for any intermediate state modulation during the crystallization process, which compromises the precise repeatability of the multi-level response. The amorphization process sees the entire cell undergoing melt-quenching with a single pulse that achieves temperatures over $T_m$, which is why multi-level with this type of heater relies mainly on tuning the nucleation during crystallization.

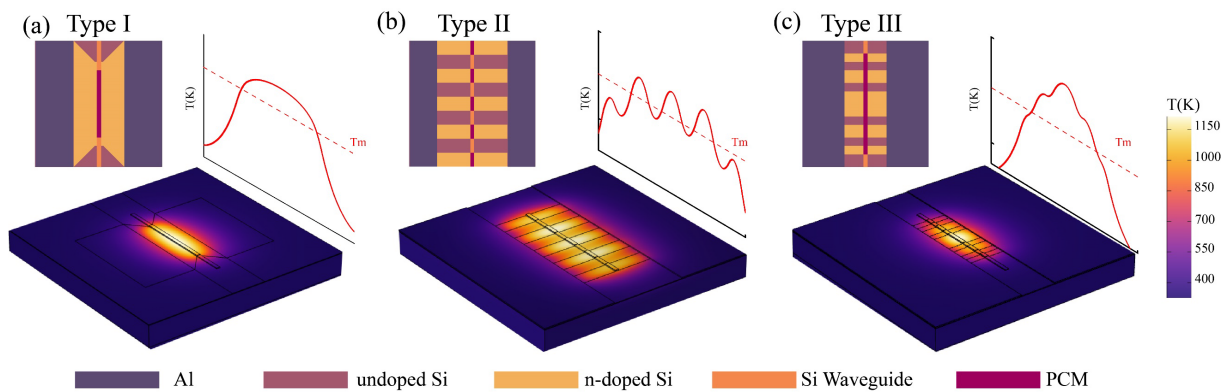

**Figure 1. Geometry and temperature profile for three types of microheaters:** (a) bowtie-shaped (b) five 10×2 μm bridges with 2 μm spacing, and (c) five 5 μm-long bridges with varying widths of 1, 1.5, 3, 1.5, 1 μm and 1 μm separation. Each figure shows the top view of the devices featuring PCM cells, silicon waveguides, and the microheater, as well as the 3D FEM temperature profiles and the temperature along the waveguide, all simulated at the end of a 400 ns and 5.5 V electrical pulse.

Fig. 1(b) shows a microheater featuring five 10 × 2 μm² bridges with 2 μm spacing between them, first demonstrated in Ref. [27] to control the amorphous/crystalline spatial distribution via pulse energy modulation. When applying a voltage pulse to the microheater, the hotspot triggers a phase transition that expands from the center to the sides of the PCM directly on top of each bridge, with a final switching area that depends on the pulse energy. By controlling the hotspot size, partial crystallization or amorphization can be achieved, leading to a multi-level optical response.[32] However, each bridge reaches similar temperatures simultaneously, limiting the number of intermediate states, i.e., the total number of achievable levels. One advantage of this device is that, during reversible cycling, the unswitched PCM between bridges remain in crystalline state. These crystalline domains can, in turn, act as nucleation seeds, ensuring efficient and complete recrystallization. This approach could overcome the device cyclability issues faced in bowtie/square heaters due to the nucleation rate reduction after repeated melt-quenching.

To further improve the temperature profile of the heaters and achieve deterministic continuous multi-level response, we propose the Type III microheater, featuring five bridges with varying widths, shown in Fig. 1(c). The length of the bridges is 5 μm while the width of the heaters varies from the center to the sides, which, from one end to the other, follows the sequence: 1, 1.5, 3, 1.5, 1 μm. This configuration achieves a triangle-like temperature profile along the waveguide, where the center will reach the highest temperature. With this microheater geometry, the phase transition starts from the center of the PCM and continuously spreads out as a function of the electrical pulse energy. Hence, careful manipulation of the pulse energy results in finer spatial variations of amorphous domains if the PCM cell is initially prepared in a crystalline state or vice versa, thus providing finer tunability of intermediate levels. In the following sections, we will demonstrate both Type II and Type III microheaters as platforms for reliable multi-level switching of waveguide-integrated PCM cells.

*Microheater Type I*

We start by showing that a doped-silicon microheater with a bowtie geometry can reach multi-level response, although only for a limited number of cycles. To do so, we used the device shown in Fig. 2(a), consisting of a ring resonator with a Type I microheater with ~$4 \times 10^{18}$ cm$^{-3}$ n-doping in the central region. We deposited 30 nm-thick and 6 µm-long $Sb_2Se_3$ and prepared it in the crystalline state (see *Materials and Methods* for more details). In this device, we achieved the lowest energy amorphization with 21 V – 400 ns pulses, following the same experimental protocols described in Ref. [26]. One single pulse was sufficient to achieve full amorphization, as expected from our analysis above since the entire PCM cell is subjected to the same temperature given the flat hotspot profile. Thus, all or most of the PCM cell reaches the melting temperature simultaneously. Interestingly, during the first crystallization cycle, we partially crystallized with 3.4 V – 0.1 ms pulses, achieving an initial multi-level response, one level per (identical) pulse. The cell was subsequently re-amorphized following a second 21 V – 400 ns pulse. Next, when attempting to achieve the same partial crystallization with 3.4 V – 0.1 ms, we only observed a small redshift, and then the device was not responsive to the same pulse. Only a 3.4 V – 1 ms managed to recrystallize fully. From this cycle onwards, the multi-level was no longer achievable, and only full amorphization (with remarkable level reproducibility) and full crystallization could be attained with a single pulse. However, the total phase shift upon crystallization reduced over time, which can be due to fewer nuclei surviving after melt-quenching and thus hindering the crystallization of specific areas, a phenomenon ultimately leading to device failure.[22] We hypothesize that the initial conditioning behavior occurs due to larger crystal domains resulting from hotplate annealing in the as-prepared state[38] vs. finer domains/grains resulting from electro-thermally crystallizing PCM with the microheater. Smaller crystal domains (and subsequent short- and mid-range ordering in amorphous) can lead to faster local crystallization since nucleation is predominant. Therefore, a single pulse that elevates the entire cell over the crystallization temperature suffices to recrystallize most domains simultaneously. We observed this behavior in other similar devices, with the multi-level disappearing within 2-3 full cycles.

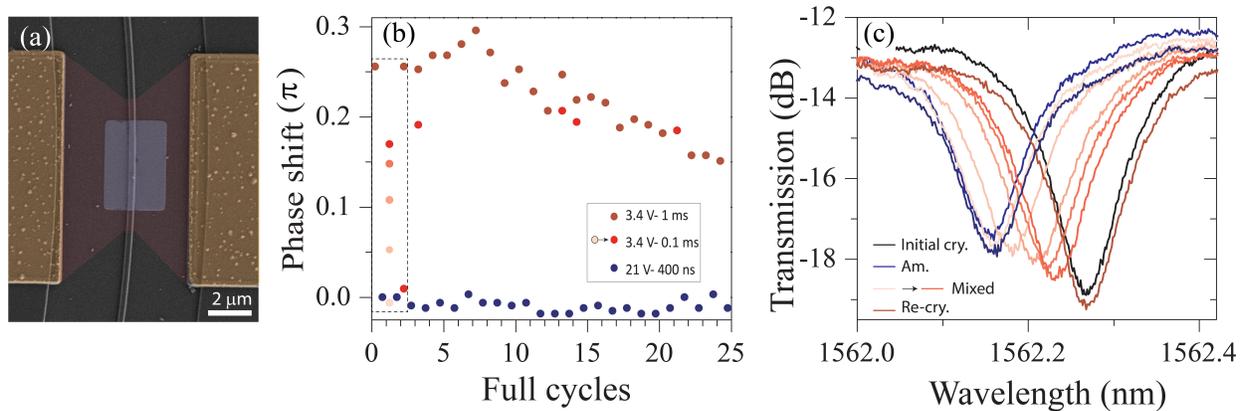

**Figure 2.** *Type I microheater*: (a) bowtie-shaped microheater with 30 nm-thick $Sb_2Se_3$. (b) Reversible switching using 3.4V - 0.1 ms to partially crystallize, 3.4V - 1 ms to fully crystallize, and 21V – 400 ns to amorphize. (c) spectra for a ring resonance showing the cycles highlighted in (b).

*Microheater Type II*

Fig. 3(a) shows a scanning electron microscope (SEM) image of a Type II microheater. This device was partially fabricated in a CMOS foundry (microheaters and metallization), followed by in-house waveguide patterning and PCM deposition. Fig. 3(b) shows the phase shifts resulting from multi-level switching of 30 nm-thick $Sb_2Se_3$ on a 120 μm-radius ring resonator for over 25 complete cycles, totaling nearly 200 switching events in three measurements over eight days. We used 500 ns pulses with voltages between 12V to 12.96 V for the five first amorphization levels and a 12.96 V - 600 ns pulse for the most extensive phase shift (i.e. maximum amorphization). Note that the significantly lower amorphization voltage in comparison to that in Fig. 2 is due to using higher n-doping for the bridges (~$10^{19}$ cm$^{-3}$ instead of ~$4 \times 10^{18}$ cm$^{-3}$ in microheater Type I shown in Fig. 2(a)—see more details in *Materials and Methods*). We added another intermediate state from the 21$^{st}$ cycle by applying a 12.96 V - 550 ns pulse, demonstrating the modulation through pulse power and width, allowing fine-tuning to reach more levels. Conversely, we achieved recrystallization using 4.32 V - 250 µs and 4.8 V - 250 µs pulses. We observe substantial variations in attaining the levels during the first three cycles due to material conditioning, followed by transient multi-levels before stabilizing at a maximum phase shift ~0.23π around cycle 13. This result indicates that less than a third of the PCM cell volume is switching reversibly based on the simulation result for the phase modulation, shown in Supplementary Fig. S4(d). The first two levels merged after 19 switching cycles, indicating that parts of the PCM can no longer switch reversibly. Interestingly, multi-level response is achieved in both switching directions. Moreover, the crystalline state displayed better stability than the amorphous state in the Type II microheater, which contradicts the results in the Type I. Since the hotspot profile depends on the pulse parameters, any variation in the high-voltage, short-length pulse to amorphize can lead to more pronounced optical response changes (also including ablation) than the crystallization pulses, which are slow enough to allow most of the domains to grow back to their full crystalline state.

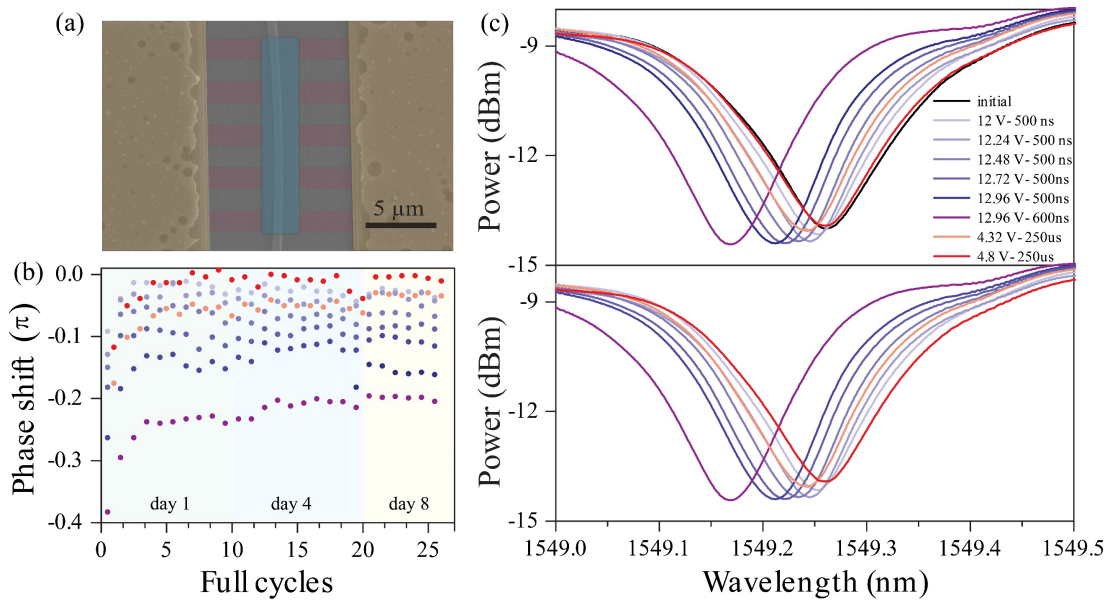

**Figure 3. Type II microheater.** (a) colored SEM picture of five-10 × 2 μm²-bridge-heater with 112.5μm radius ring resonator (b) multi-level phase shift of a resonance as result of switching $Sb_2Se_3$ reversibly, the amorphization and crystallization steps follow the same color scheme as in (c) and (d). (c) Optical spectrum measurement results in two cycles switching with 6 levels from the 13th (top) and 14th cycle (bottom). The initial state corresponds to the hotplate fully crystallized state.

Fig. 3(c) shows the optical spectrum of two consecutive cycles. As expected, the resonance dip shifted gradually to shorter wavelengths as the amorphous domain increased in area within the PCM cell. The extinction ratio increased from 5.58 dB in the initial state to 6.46 dB in the most amorphous state, which we attribute to variations in scattering, which increase with the highest refractive index state. A comparison between the initial state (fully crystallized on a hotplate after deposition) and the recrystallized spectrum in Fig. 3(c) reveals no crystalline resonance peak shift, suggesting no damage after cycling.

*Microheater Type III*

We first discuss devices comprising n-doped Type III microheaters with no waveguides, as shown in Fig. 4(a), in order to characterize their thermal response via transient thermoreflectance imaging.[39] Fig. 4(b) shows the simulated temperature profiles under 400 ns voltage pulses. The temperature ratio between the central 3µm-width bridge and the 1µm-width bridges on the sides increases with voltage. As a result, the overall temperature profile changes from a mostly flat at low temperatures (i.e., triggering uniform crystallization) to a triangle-like profile when reaching the melting point (i.e., triggering differential amorphization). Fig. 4(c) shows the experimental temperature distribution of the fabricated device using the pixel by pixel thermoreflectance coefficient shown in Fig. 4(a)[40]. Given the complexity of mapping accurately the temperature when using 400 ns probing, the maximum reliable temperature measured was 800 K; however, as discussed later, the heater can be driven at higher voltages, reaching temperatures over the melting point. At 10 V the heater reaches the crystallization temperature with 400 ns pulses. However, the temperature at the center of the 3-μm bridge can also be controlled by changing the pulse width. As we show in Supplementary Fig. S1, the surface temperature of the heater reaches 80% of its maximum temperature after approximately 800 ns and reaches a steady state for pulses longer than 3 μs.

Similar to the simulations, the experimental results in Fig. 4(c) show that—at the maximum temperature we could probe—there is only a 15% maximum difference between the center-to-edge temperature ratio (measured at positions located at 9 μm and 3 μm in Fig. 4(b)), which indicates that the triangle-like behavior in our devices will be reached at higher temperatures if compared to the simulations. Additionally, the experimental profiles display temperature dips between bridges that we do not observe in Fig. 4(b). To further study these effects, we performed simulations with the actual size of the metal pads, as opposed to the initial simplified device shown in Fig. 1. We found that heat dissipates towards the metal electrodes faster than to the undoped Si and substrate, given their significantly larger size (>11000 μm$^2$), which our design did not account for initially. We found that both the dips between bridges and the fast dissipation on either side of the microheater are caused by this faster dissipation, as demonstrated in Supplementary Fig. S2. The temperature-dependent thermal conductivity coefficient for thin film silicon should also be considered for a more accurate simulation and design since heat spreading will be decelerated at a higher temperature or in thinner Si films.[41,42]

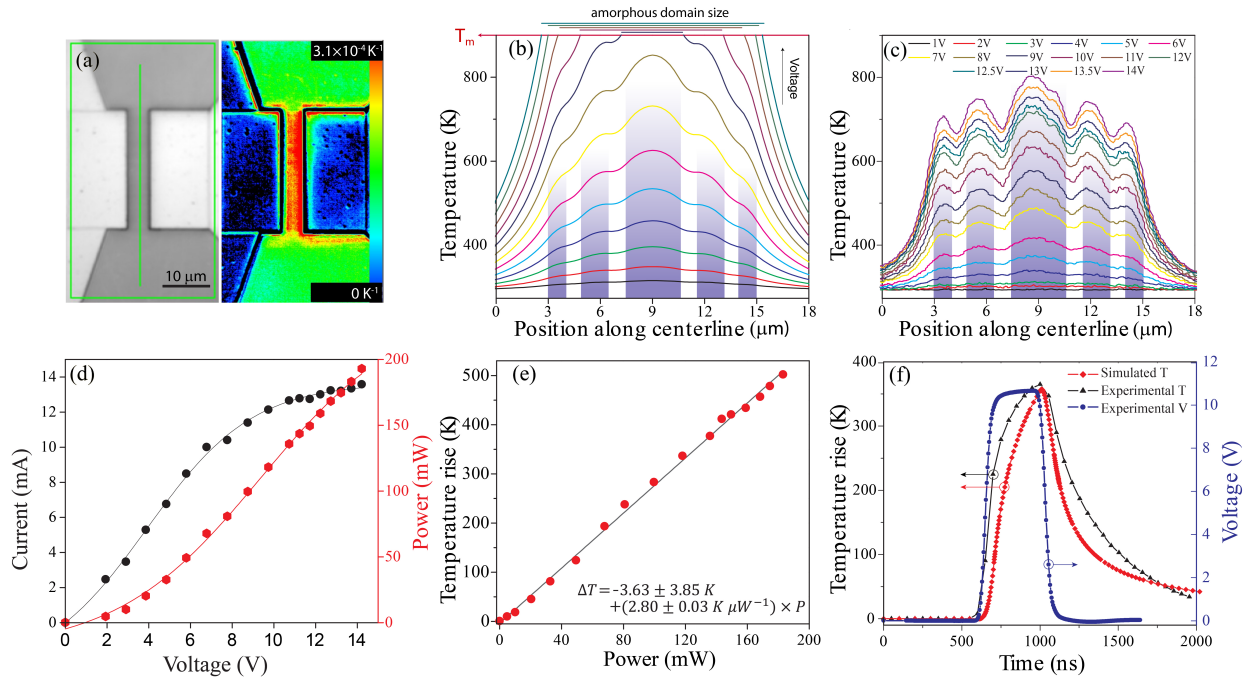

**Figure 4. Thermoreflectance characterization of microheater Type III without waveguide.** (a) Optical microscope image of the device (left) and mapping of the thermoreflectance coefficients using a 780 nm illumination source (right). (b-c) The temperature vs position along the centerline of the bridges from (b) 3D-FEM simulation after applying 400 ns pulses with increasing voltages and (c) experimental results using voltages ranging from 1 V to 14 V (maximum reliable voltage for thermoreflectance measurements). The shaded mark the location of the five bridges. (d) Current and power as a function of the voltage applied on the microheaters. (e) Temperature rise from room temperature as a function of the pulse power. Temperature is averaged over the length of the microheater. (f) Experimental voltage pulse (11 V - 400 ns) and experimental and simulated real-time temperature at the center point of the central bridge.

Fig. 4(d) shows the average applied current and dissipated power for each 400 ns pulse as a function of voltage for the multi-width bridge microheater without a waveguide. Based on the current-voltage (IV) curve, the device exhibits constant resistance (550~600 Ω) below 7 V applied voltage. Once the voltage applied is larger than 7 V, the slope of the curve reduces leading to 900~1200 Ω resistance and demonstrating a nonlinear response typical of single-doped silicon heaters.[26] Under the thermoreflectance setup, we also found that a 14.5 V (on the device) pulse will damage the heater by melting silicon.[39] In Fig. 4(e), we further show that the mean temperature along the width of the microheater is linear with voltage, indicating that the efficiency of Joule heating is stable within the operation range.

We then analyze the real-time heating and cooling response of our devices using 11 V-400 ns pulses featuring rise and fall times of 95 ns and 80 ns, respectively. Fig. 4(f) shows that this pulse elevates the temperature by 365 K to achieve ~658 K, capable of crystallizing most PCMs. The experimental results display slightly faster heating, but slower cooling rates than the simulated results, indicating that the simulations overestimate the overall thermal diffusivity. For instance, during the cooling stage, the experimental device and the model took 115 ns and 180 ns, respectively, to cool from the peak to the crystallization temperature, suggesting a slower yet acceptable cooling rate for the reamorphization of most PCMs. This discrepancy is due to the thermal mass of the device, which is simplified and idealized in the simulation.

To fully visualize the temperature profiles shown in Fig. 4(c), we recorded thermal images at the end of each 400 ns pulse for several voltages. The results are shown in Fig. 5(a), where we observe Joule heating mostly confined within the range of the microheater. The temperature gradient along the bridge is uniform within 1 μm around the central position, thus ensuring uniform heating of the PCM in this direction and leaving the spatial temperature modulation exclusively to the direction perpendicular to the bridges. We notice that the temperature profile is not perfectly centered, and the highest average temperature is shifted to the right by 0.6 µm. This rightward shift can be attributed to the directional current flow (forward or reverse bias) which constructed the carrier distribution along the junction. The volume difference between the two metal electrodes we used in our experiment, where the side with the larger electrode (left) undergoes faster heat dissipation, also contributes to the shifting.

We now demonstrate the partial amorphization of PCMs using color variations of 1 µm-wide, 10 µm-long GSST cell under optical microscopy. The result of intermediate amorphization and the pulses used are shown in Fig. 5(b). The phase transformation was confirmed using Raman spectroscopy, as shown in Fig. 5(c), where a Raman peak at 120 cm$^{-1}$ is found only in crystalline GSST.[24] To perform this experiment, the sputtered amorphous GSST was first annealed at 325 °C for 30 minutes to crystallize the cell fully. We found that visible color variations denoting amorphization take place for voltages higher than 15 V. Note that this voltage is for our custom MOSFET amplifier system, as discussed in the Methods section, which differs from those used in thermoreflectance measurements (likely due to differences in impedance mismatch between the device and different pulse generators). The latter technique accurately measures the voltage dropped on the device (using a high speed pulsed IV system), while the former corresponds to the DC voltage applied to the MOSFET amplifier. By increasing the voltage of the amplifier, we confirmed that the amorphization domain grows controllably as a function of the pulse voltage/energy until breaking at around 16.5 V. The GSST cells switched with 15.5 V, 16 V, and 16.3 V displayed lengths of 2.7 μm, 8.1 μm, and 8.5 μm amorphous domains embedded within the initially prepared crystalline cell. To fully recrystallize GSST, we applied a pulse of 6V for 400 ms. The Raman analysis on the different states confirms the occurrence of both switching events, thus confirming the capability of the multi-width heater in step amorphization and single-step crystallization.

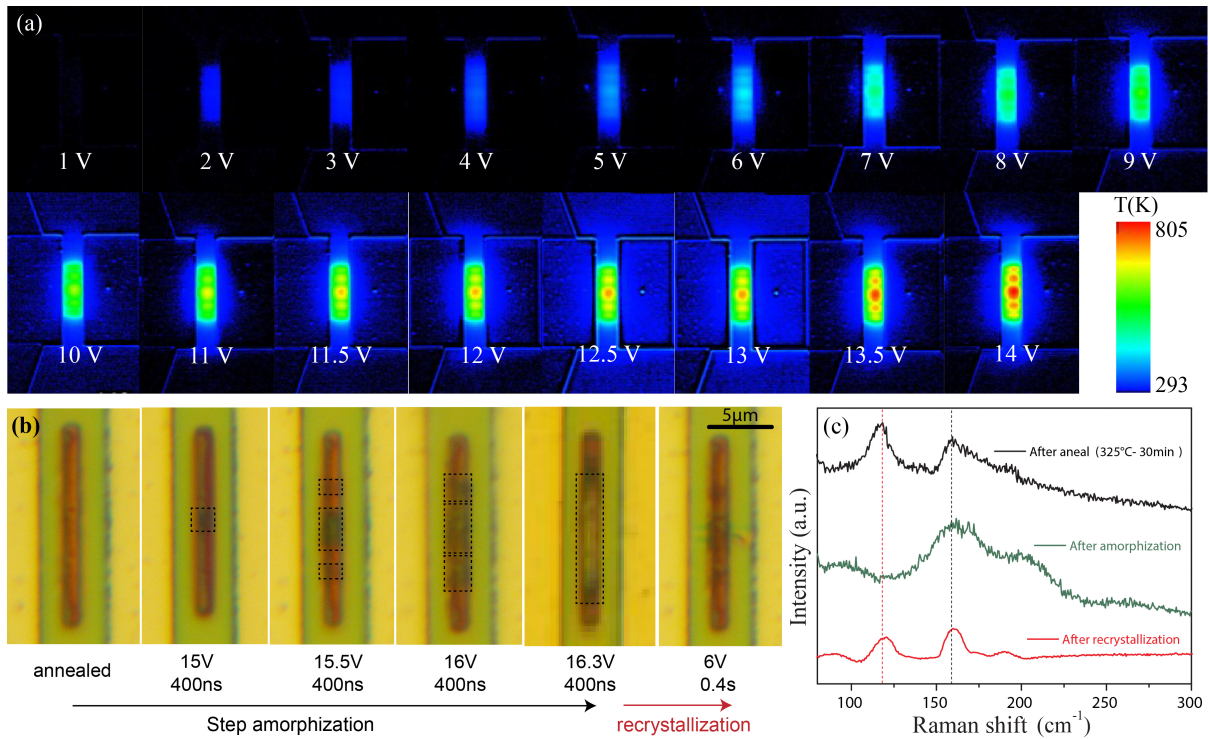

**Figure 5. Transient thermoreflectance and partial switching measurements in Type III microheaters.** (a) Transient thermoreflectance images at the end of 400ns pulses with varying voltages applied to the device. (b) Optical microscope images taken for the initial annealed crystalline GSST and the subsequent step amorphizations and final recrystallization. The scale bar corresponds to 4 µm. (c) Raman spectra measured at the center of the PCM cell after annealing (black); after a 16.3V-400ns pulse (green); and after a 6V-400ms pulse (red). We note that the images were all taken in the same device except for the 16.3V, which corresponds to a second identical device that was switched following smaller voltage steps.

Furthermore, we patterned photonic integrated waveguides within n-doped Type III microheaters for reliable thermoreflectance characterization (the oxide cladding on commercial foundry samples could introduce thin film interference that can result in inaccurate measurements.[43]) Fig. 6(a) shows the SEM picture of an 80 μm-radius ring resonator on a 150 nm SOI platform with a 30 μm-thick $Sb_2Se_3$ cell deposited on the waveguide. Due to the etching required to pattern a silicon waveguide, the device's resistance is higher than the device without a waveguide, thus requiring a larger voltage to achieve the same temperature. According to our simulations in Supplementary Fig. S4 and assuming ideal ohmic behavior, a 76% power increase is required to achieve the same temperature when a waveguide is patterned within the heater. In Fig. 6(b), we track the temperatures at three different points (labelled in Fig. 6(a)) on the central horizontal line of the heater. The heater's middle point coincides with the waveguide and is separated by 1μm from the other reference points on either side. The waveguide shows the highest temperature, while the point on the inner side of the ring has a higher temperature than the one on the outer side, which we once again attribute to dissimilar heat dissipation due to different size metal pads on each side. The 2D temperature map for the top view at the end of the pulse is shown in Fig.6(c). The results demonstrate the expected heat confinement on the waveguide, and in contrast to the results in Fig. 5, the patterning of the waveguide leads to a closer experimental triangle-like temperature profile than that simulated. Transient thermoreflectance characterization with various pulses in this device is included in Supplemental Fig. S5. Lastly, Fig. 6(e) shows the real-time temperature response at the center of the device in Fig. 6(c) using 16 V-500 ns pulses, enough to raise the temperature by 150 K. The results show that a device with a waveguide displays a lower cooling rate, which we attribute to the poor thermal conductivity of etched-silicon trenches that prevent heat from dissipating towards the electrodes.[42]

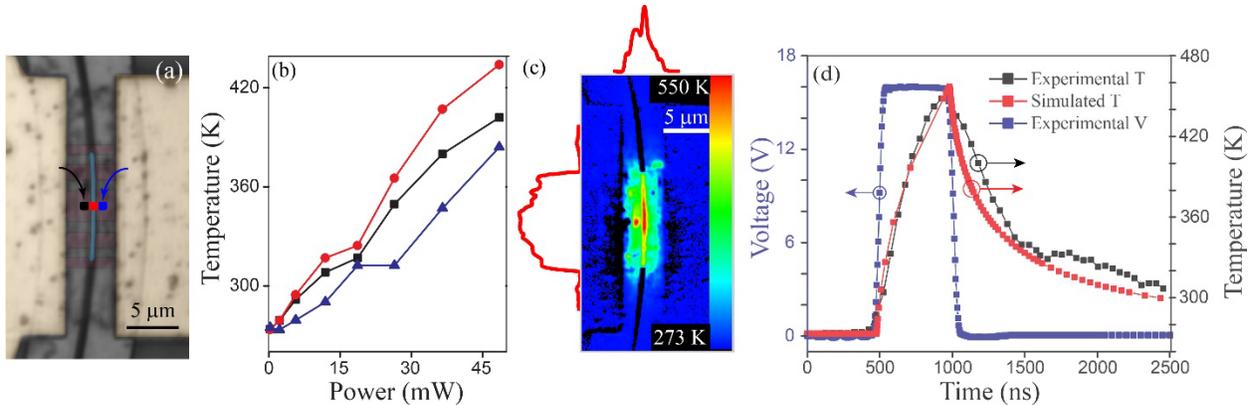

**Fig 6**. **Type III microheater with a waveguide for thermoreflectance characterization**. (a) SEM image of the device with three temperature-tracking points (b) Maximum temperature of the tracking points as a function of electrical power applied on the device. (c)&(d) Transient thermoreflectance images of 2 devices at the end of 500 ns pulse with (c) 16 V (d) 18 V (e) The applied voltage pulse (16 V-500 ns) and cooling curve in an experiment comparing with the simulation cooling curve.

*Microheater Type III in zero-change commercial foundry-processed PICs*

Lastly, we demonstrate Type III microheaters using n-doped and p-doped silicon and $Sb_2Se_3$ and GSST cells embedded in Mach-Zehnder interferometers (MZIs). Fig.7(a) shows optical microscope images of the PICs taped out by Advanced Micro Foundry (AMF), featuring microheaters on either arm of the MZIs for loss balancing but only one with external electrical control. Our devices included the oxide cladding etching layer offered by AMF, which we use to pattern windows on the microheaters and deposit PCMs in direct contact with the waveguide (see Methods). Fig. 7(b) shows a schematic of the cross-section of the final device, including a final oxide deposited to cap the PCM cell. We measured a resistance of 420-450 Ω for n-doped heaters and a significantly larger 5.5-6 kΩ for the p-doped. All PCMs remained in the amorphous state before the testing. Fig. 7(c) shows the multi-level switching of 30 nm-thick GSST using an n-doped Type III microheater. We plot the spectra of the multiple levels involved in a single re-amorphization process but show the results of complete reversible cycles in the inset. We note that amorphization increases the optical path length by reducing the $n_{eff}$ on the shorter MZI arm, thus leading to a redshift. We used 10 μs - 5.28V to 7.32V pulse to achieve area-selective melt-quenching and, thus, partial amorphization, and 10 ms - 4.32V pulses to crystallize, with a maximum observed phase shift of approximately π/5 and a total extinction ratio variation of ~20 dB. Because crystalline GSST is absorptive and amorphous GSST is transparent, we also observe a modulation in the visibility of the interferogram by ~4 dB due to the imbalance in losses between both arms (the GSST on the reference arm remained in the amorphous state throughout all the measurements). Fig. 7(d) displays the transmission spectra of the device with a p-doped heater and a 30 nm-thick $Sb_2Se_3$ cell measured in one amorphization cycle. We achieve a similar ~π/5 phase shift with 50 μs - 12.48 V pulses (a higher voltage was required to compensate for the more considerable resistance observed in these heaters), and 500ns – 28 V with 0.5 V increments to achieve intermediate states through partial amorphization.

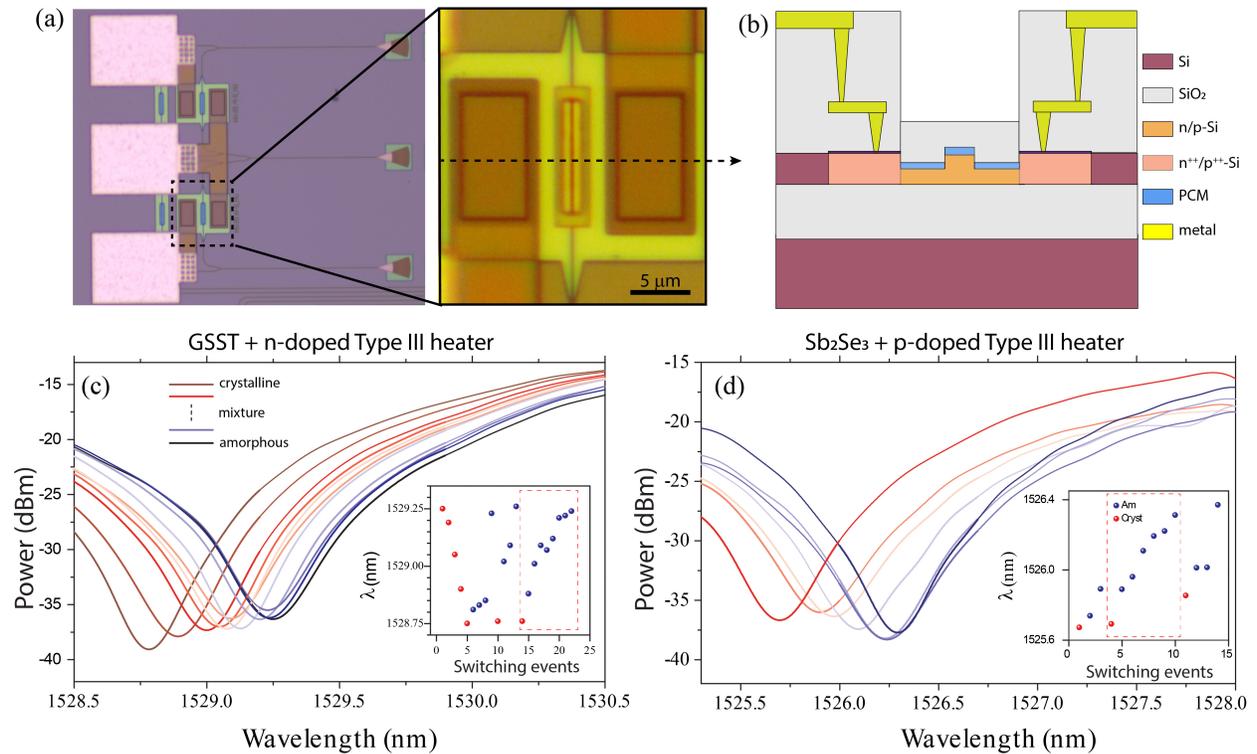

**Figure 7. Commercial foundry-processed PICs with Type III microheater**. (a) Optical microscope images of two unbalanced MZIs and zoom-in to a single Type III microheater with embedded PCM. (b) Illustration of an approximate crosssection (not in scale). (c) and (d) Zoom-in to the MZIs' interferograms tracking the modulation at a minimum in transmission using (b) 30-nm thick GSST on a n-doped and (c) 30-nm thick $Sb_2Se_3$ on a p-doped Type III microheater.

## Conclusion

We have proposed and experimentally demonstrated doped-silicon microheaters with engineered temperature profiles to deterministically control partial amorphization via electrical pulse energy modulation. With our approach, the multi-level response of photonic devices would rely on the deterministic length of an amorphous domain embedded in a crystalline cell rather than on the stochastic nucleation during the crystallization process. We demonstrated the versatility of our approach by applying the multi-level concept in two novel geometries with n-doped and p-doped silicon as conductive materials, with GSST and $Sb_2Se_3$ PCM cells, and in devices following two instances of CMOS fabrications: partial foundry with in-house processing and zero-change commercial foundry with BEOL deposition of PCM. We then used such microheaters to demonstrate active, multi-level, nonvolatile phase and amplitude modulation with $Sb_2Se_3$ and GSST in ultra-compact, electrically-driven silicon photonic devices embedded in ring resonators and MZIs.

Furthermore, we fabricated the microheaters based on simulation models as the guiding design; however, we found other parameters that impacted the experimental response of our heater. In particular, the size of the metal pad for electrical contacts is a determining factor of the heat dissipation and, thus, of the temperature profiles. While the current designs—especially those with in-house fabrication in Fig. 5—still perform multi-level amorphization, they can be further optimized to compensate for these effects. Topology optimization methods can be an exciting route to inform novel and improved geometries to engineer hotspots.[44] Moreover, we demonstrated computationally and experimentally that when etching silicon to form a waveguide, the gradient of the temperature profile is optimized, enabling finer control on the amorphous domain switching; however, thinner silicon slabs come at the cost of larger resistances, thus requiring higher switching energies.

Our work provides insight into engineering the Joule heating in microheaters by simply controlling its topology. Future studies could focus on improving temperature distribution, minimizing defects in electrical and thermal performance, and switching and testing the reproducibility of the multi-step amorphization of PCMs on PICs with higher phase shifts. We also note that the geometry optimization strategy can be applied to other microheater platforms such as

PIN doped-silicon,[28] graphene,[24,25] and conductive oxide[32] and expanded to microheaters controlling 2D temperature profiles for metasurfaces.[16]

## Materials and methods

*Fabrication*

The samples in Fig. 2 and Fig. 3-6 were fabricated on 220-nm silicon-on-insulator (SOI) wafers with phosphorous n (~4 × $10^{18}$ $cm^{-3}$) and (~$10^{19}$ $cm^{-3}$), respectively. The contacts for all these devices were formed with $n^{++}$ (~$10^{20}$ $cm^{-3}$) doped regions patterned via ion-implantation and a layer of 100 nm aluminum on top of these regions to enhance the contact and electrical conductivity between the microheater and the electrical probes. These devices followed the fabrication process described by Ríos *et al*[27]. Two $n^{++}$ regions are connected via an $n^+$ doped bowtie or bridges to concentrate the heat while allowing for silicon waveguides to guide light with low optical losses.[27]. A 30-nm-thick $Sb_2Se_3$ or GSST thin-film was deposited onto the microheaters using an AJA Orion-3 Ultra High Vacuum Sputtering system at room temperature. Patterning of the PCM cells was performed using electron beam lithography on an Elionix ELS-G100 system with Ma–N 2403 negative resist, followed by $CF_4$ reactive-ion etching. 30 nm of $Al_2O_3$ was subsequently deposited via Atomic Layer Deposition (ALD) to protect the PCM from oxidation.

The devices in Fig. 6 were fabricated at the Advanced Micro Foundry in Singapore, using their silicon-on-insulator process. Open oxide windows through the top cladding layer was used to access the waveguide for PCM deposition. The PCM was deposited and patterned using the same method described above followed by capping with 30 nm $SiO_2$.

*Device simulation:*

The transient temperature profile of doped-silicon microheaters, under pulsed electrical biasing, was investigated using time-dependent three-dimensional finite-element-method (3D-FEM) simulations with COMSOL Multiphysics®. The doped regions' electrical properties of the devices were simulated using the Semiconductor module using the material properties listed in Table 1. Voltage pulses were applied on the Al pads, and the substrate was assumed to be an infinite volume with a stable temperature at 293.15 K. Constant thermal properties were assumed for both GSST phase states, given their small values and thermal mass when compared to the other dominant elements in the device.[45] The optical simulations shown in the Supplementary S4 were performed using Lumerical FDTD (Ansys®)

Table 1. Material properties used in simulations

|  | Density (kg/m$^3$) | Specific heat (J/(kg·K)) | Thermal conductivity (W/(m·K)) |
|---|---|---|---|
| **Si** | 2329 | 700 | 131 |
| **SiO$_2$** | 2203 | 730 | 1.4 |
| **GSST** | 6140 | 293[46] | 0.48[46] |
| **Al** | 2700 | 890 | 230 |

***Thermoreflectance characterization:***

Transient thermoreflectance imaging (TTI) was used to characterize the microheater's transient thermal dynamics[47]. Based on a CCD approach, a single wavelength LED (780 nm) was pulsed to measure the surface reflectance change of the microheater. This method employs a lock-in averaging approach that allows for simultaneous high spatial resolution (≈ 150 nm/pixel) and temporal resolution (≈ 50 ns)[40]. When attempting to directly probe the reflectance of a semiconductor, the thermoreflectance signal can be significantly improved by using excitation wavelengths near/above the bandgap of the semiconductor[48]. In this case, a visible wavelength LED (≈ 1.59 eV) was used to probe the surface temperature rise of the SOI microheater and a LED (≈ 1.88 eV) was used to probe the PCM on the heater with waveguides.

The accuracy of TTI relies on determining the thermoreflectance coefficient, $C_{TH}$ (assuming a linear relationship between temperature rise and reflectance change). The $C_{TH}$ of the top layer (silicon) was experimentally found using a 100× objective (NA = 0.7) and a temperature-controlled stage. The change in thermoreflectance ($\Delta R/R$) was measured for a given set of temperature rises by increasing the stage temperature from 20°C to 120 °C (in 20 °C steps). The surface temperature rise is independently measured with a thermocouple that is positioned near the device with thermal paste. Using an iterative approach, the $C_{TH}$ of the heater region was monitored and calibrations were repeated until the $C_{TH}$ value converged. Averaging over multiple pixels, the standard deviation, with 95% confidence intervals, was used to estimate the $C_{TH}$ uncertainty. For the 780 nm and 625 nm excitation, the $C_{TH}$ of the silicon microheater and for the PCM was measured to be $(2.9 \pm 0.15) \times 10^{-4}$ K$^{-1}$ and $(3.36 \pm 0.12) \times 10^{-4}$ K$^{-1}$ respectively. A pulsed IV system was used to electrically bias the microheater (400 ns pulse) for TTI. Based on a lock-in approach, an internal trigger was used to synchronize the pulsed IV with the TTI system. A 10% duty cycle was implemented to ensure that the heater has sufficient time to cool and return to its initial temperature after each pulse. The transient temperature represents the average temperature of the region of interest (ROI) defined at the center of the microheater (green box in the inset of Fig. 3(d)). The ROI captures the region in the microheater that exhibits the highest temperature.

*Optical and Electrical measurements:*

A custom-made wafer-scale photonic testing system was used to perform the optoelectronic measurements. A custom 10-channel SMF-28 Ultra Fiber Array with a 127 μm pitch (Innoall®) coupled light in and out of the chip using silicon-etched grating couplers. A Santec TSL-570 with a Santec MPM-211 photodetector setup was used to measure the transmission spectrum of the PICs. The electrical switching of PCM was achieved using a Moku:Pro (Liquid Instruments®) as a pulse generator. To overcome the maximum voltage limitations of the Moku:Pro, a digital output line was connected to an integrated GaN-FET half bridge chip (MASTERGAN1 by STMicro) to amplify rectangular pulses for amorphization (joule heating to melting point and quench), and an analog output line was connected to a wide bandwidth current feedback amplifier (THS3491DDA by Texas Instruments) to amplify arbitrarily definable pulses for crystallization (low-speed joule heating and keep for longer times). A Keithley 2400 and a Rockseed RS305D were used as DC power sources.

*Raman characterization:*

The PCM phase identification was performed using surface-enhanced Raman spectroscopy (Yvon Jobin LabRam ARAMIS) with a 532 nm laser, a 2400 lines/mm grating, and a 100× long-working-distance objective.

# Acknowledgements

This work has been supported by ONR (MURI N00014-17-1-2661), AFOSR (FA9550-24-1-0064), the National Science Foundation (ECCS-2210168/2210169, ECCS-2132929, CISE-2105972, CCSS-2337674, and DMR-2329087/2329088) and is supported in part by funds from federal agency and industry partners as specified in the Future of Semiconductors (FuSe) program. C.R.O. acknowledges support from the Minta Martin Foundation through the University of Maryland.